\documentclass[twocolumn,aps,prl,amsmath,amssymb,superscriptaddress]{revtex4-1}

\usepackage{mathrsfs}
\usepackage{graphicx}
\usepackage{times}

\newcommand{\up}{\mid\uparrow\rangle}
\newcommand{\dn}{\mid\downarrow\rangle}
\newcommand{\probe}{\mid\!\!\mathrm{p}\rangle}
\newcommand{\ket}[1]{\mid\!\!#1\rangle}

\begin{document} 
\title{Transverse Demagnetization Dynamics of a Unitary Fermi Gas} 
\author{A.\ B.\ Bardon}
\author{S.\ Beattie}
\author{C.\ Luciuk}
\author{W.\ Cairncross}
\author{D.\ Fine}
\author{N.\ S.\ Cheng}
\author{G.\ J.\ A.\ Edge}
\affiliation{Department of Physics and CQIQC, University of Toronto, M5S 1A7 Canada}
\author{E.\ Taylor}
\affiliation{Department of Physics and Astronomy, McMaster University, L8S 4M1 Canada}
\author{S.\ Zhang}
\affiliation{Department of Physics, Center of Theoretical and Computational Physics, University of Hong Kong, China}
\author{S.\ Trotzky}
\author{J.\ H.\ Thywissen}
\affiliation{Department of Physics and CQIQC, University of Toronto, M5S 1A7 Canada}
\affiliation{Canadian Institute for Advanced Research, Toronto, Ontario, M5G 1Z8 Canada}
\begin{abstract}
Understanding the quantum dynamics of strongly interacting fermions is a problem relevant to diverse forms of matter, including high-temperature superconductors, neutron stars, and quark-gluon plasma. An appealing benchmark is offered by cold atomic gases in the unitary limit of strong interactions. Here we study the dynamics of a transversely magnetized unitary Fermi gas in an inhomogeneous magnetic field. We observe the demagnetization of the gas, caused by diffusive spin transport. At low temperatures, the diffusion constant saturates to the conjectured quantum-mechanical lower bound $\simeq \hbar/m$, where $m$ is the particle mass. The development of pair correlations, indicating the transformation of the initially non-interacting gas towards a unitary spin mixture, is observed by measuring Tan's contact parameter.
\end{abstract}
\maketitle

Short-range interactions reach their quantum-mechanical limit when the scattering length that characterizes inter-particle collisions diverges. A well controlled model system that realizes this {unitary} regime is provided by ultracold fermionic alkali atoms tuned to a Fano-Feshbach resonance \cite{Review2012}. These scale-invariant gases are characterized by universal parameters relevant to diverse systems such as the crust of neutron stars at twenty-five orders of magnitude higher density \cite{Baker:1999,Ho:2004gm}. Experiments with ultracold atoms have already greatly contributed to the understanding of equilibrium properties of unitary gases \cite{Nascimbene:2010en,Horikoshi:2010,Ku:2012ue}. Progress has also been made in the study of unitary dynamics \cite{Thomas2010,Hulet2011,Zwierlein2011,Kohl2013,Sidorenkov:2013hk}, including observations of suppressed momentum transport \cite{Thomas2010} and spin transport\cite{Hulet2011,Zwierlein2011,Kohl2013} due to strong scattering.

Spin diffusion is the transport phenomenon that relaxes magnetic inhomogeneities in a many-body system. At low temperature, where Pauli blocking suppresses collision rates, one must distinguish between diffusion driven by gradients in either the magnitude or the direction of magnetization, and quantified by longitudinal spin diffusivity $D_s^{||}$ or transverse spin diffusivity $D_s^{\perp}$, respectively \cite{Meyerovich:1985pla,Jeon:1989}.
A measurement of $D_s^{||}$ in a three-dimensional unitary Fermi gas yielded a minimum trap-averaged value of $6.3(3) \hbar/m$ \cite{Zwierlein2011}.
This is consistent with a dimensional argument, in which diffusivity is a typical velocity ($\hbar k_F/m$ for a cold Fermi gas, where $\hbar k_F$ is the Fermi momentum) times the mean free path between collisions. In the absence of localization, the mean-free path in a gas cannot be smaller than the interparticle spacing $\sim 1/k_F$, which translates into a quantum lower bound of roughly $\hbar/m$ \cite{Zwierlein2011,Levin:2011,Enss:2012b}.
However, $D_s^{\perp}$ as low as $0.0063(8) \hbar/m$ was recently observed in a strongly interacting two-dimensional Fermi gas \cite{Kohl2013}. This thousand-fold range in transport coefficients remains unexplained by theory.

\begin{figure}[b]\begin{center}
\includegraphics[width= 0.9 \columnwidth]{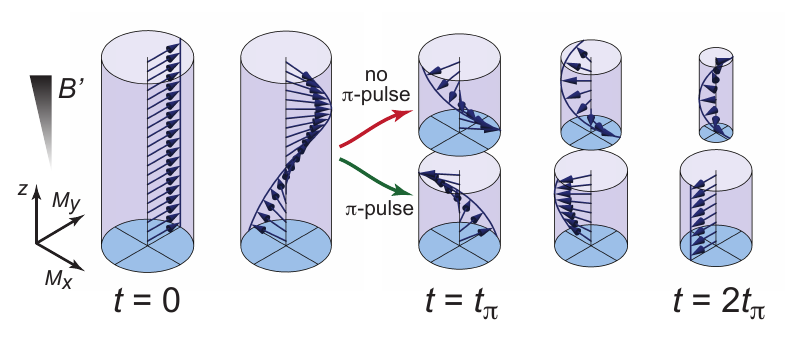}
\caption{{\bf Magnetization dynamics.} A $\pi/2$ pulse at $t=0$ initializes the system with a homogeneous magnetization ($M_y=1$ in the rotating frame) perpendicular to the magnetic field, which is along $z$. A spin spiral develops because of a magnetic field gradient, and drives diffusive spin currents. The upper and lower sequences show evolution without and with a $\pi$ pulse, respectively. \label{fig:cartoonmag}}
\end{center}\end{figure}

We measure the transverse demagnetization dynamics of a three-dimensional Fermi gas that is initially fully spin-polarized. All of our measurements are carried out with samples of ultracold $^{40}$K atoms in a harmonic trap. Each atom is prepared in an equal superposition of two resonantly interacting internal states, labeled $\up$ and $\dn$ \cite{SM}, which corresponds to a gas with full transverse magnetization $M_y=1$ (Fig.~\ref{fig:cartoonmag}). Initially, interactions between these identical ultracold fermions is inhibited by the Pauli exclusion principle. The states we use also block any local mechanism for spin relaxation, unlike the scenario typical in liquids or solids. However, the differential magnetic moment $\Delta\mu$ between the internal states allows a magnetic field gradient $B' =\partial B_z/\partial z$ to twist the magnetization across the cloud into a spiral pattern, leading to a gradient in transverse magnetization. This gradient drives diffusive spin transport that erodes the coherence irreversibly. In contrast, for a weakly interacting Fermi gas, collisionless spin waves lead to reversible dynamics \cite{Du:2009hm}. For the strongly interacting Fermi gas probed here, the evolution of transverse magnetization $M_\perp =M_x + i M_y$ is modeled with $\partial_t M_\perp = -i\alpha z M_\perp + D_s^\perp \nabla^2 M_\perp$, neglecting trap effects, where $\alpha = \Delta\mu B'/\hbar$ \cite{Abragam}. 
This equation is solved by $M_\perp(\vec r,t) = i\exp[-i\alpha z t - D_s^\perp \alpha^2 t^3/3],$ 
such that $(D_s^\perp \alpha^2)^{-1/3}$ gives the time scale of demagnetization.
Because there is no spatial gradient in the magnitude of magnetization, the dynamics do not probe $D_s^{||}$.

The effect of spin diffusion on magnetization is measured using the spin-echo technique described in Fig.~\ref{fig:cartoonmag}. The spin-refocusing $\pi$ pulse at $t_\pi$ swaps the population of the states $\up$ and $\dn$, which causes the spin spiral to start untwisting. This partial rephasing also reduces the rate of diffusion. At $t=2 t_\pi$, the model anticipates a spin echo with
\begin{equation}\label{eq:tau} 
M_\perp(t) = -i \exp[-D_s^\perp \alpha^2 t^3/12].
\end{equation}
The final cloud-averaged $|M_\perp|$ is indicated by the contrast in $\up$ and $\dn$ atom number after a final $\pi/2$ pulse with variable phase \cite{SM}.

\begin{figure}[tb]\begin{center}
\includegraphics[width=0.9 \columnwidth]{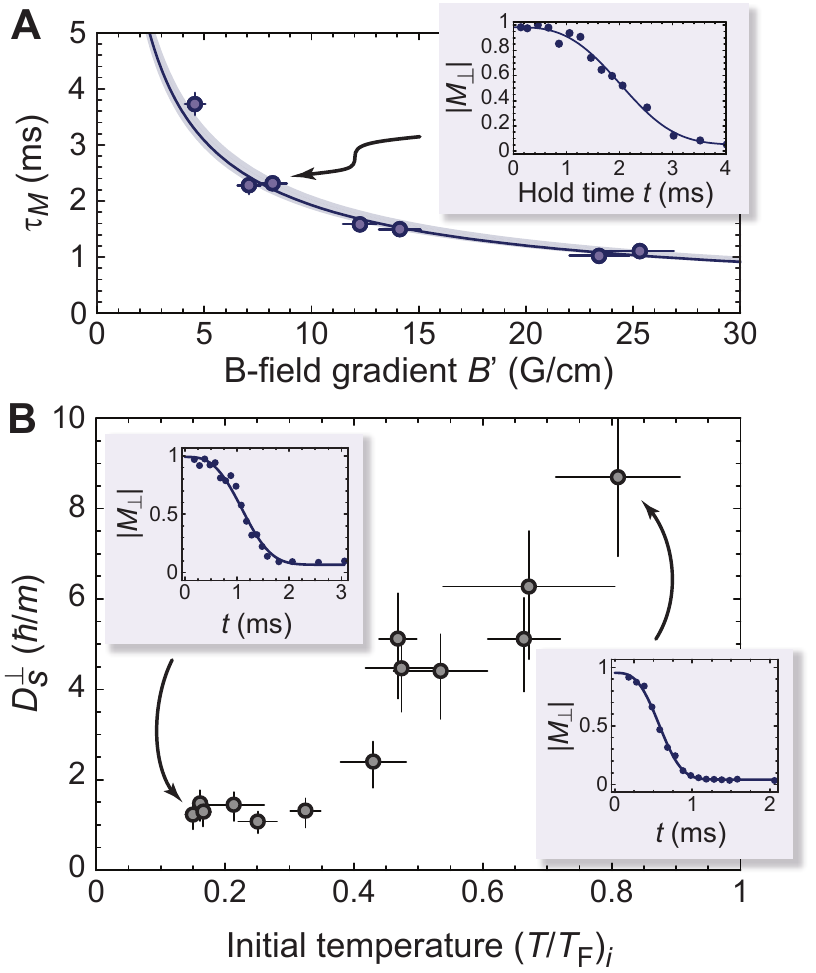}
\caption{{\bf Spin diffusion.}
{\bf (A)} Magnetization decay times $\tau_M$  at various gradients at an initial temperature of $(T/T_F)_i=0.25(3).$ Horizontal and vertical error bars reflect the systematic error in the gradient calibration and the fit error respectively. The solid line is a single-parameter fit of $\tau_M = (D_s^\perp \alpha^2/12)^{-1/3}$ to the data, and the shaded area denotes the combined statistical and systematic uncertainty.
The inset shows a sample of $|M_\perp(t)|$ at $B'=8.2(7)$\,G/cm with a fit using Eq.~\ref{eq:tau}.
{\bf (B)} Measured $D_s^\perp$ for various initial temperatures at a single gradient $B' = 18(1)$\,G/cm. To achieve temperature control, the data were collected with $N = 1-7 \times 10^4$. Horizontal error bars are statistical uncertainty, vertical error bars combine  gradient-calibration and fit uncertainty. The insets show sample data for $|M_\perp(t)|$ at two initial temperatures with a fit using Eq.~\ref{eq:tau}.
\label{fig:mag}}
\end{center}\end{figure}

We observe that demagnetization occurs in several milliseconds (Fig.~\ref{fig:mag}A, inset). Fitting $|M_\perp(t)|$ with an exponential decay function $\exp[-(t/\tau_M)^\eta]$, we find a range of $2.5 \lesssim \eta \lesssim 4.0$, compatible with $\eta=3$ in Eq.~\ref{eq:tau}.  Constraining $\eta=3$, we extract $\tau_M$ across a wide range of gradients (Fig.~\ref{fig:mag}A), and fit it to find that the $(B')^{-2/3}$ scaling of Eq.~\ref{eq:tau} holds even for the case of the trap-averaged magnetization. At an initial temperature $(T/T_F)_i = 0.25(3)$, where $T_F$ is the Fermi temperature of the spin-polarized gas \cite{SM}, a single-parameter fit of $\tau_M = (D_s^\perp \alpha^2/12)^{-1/3}$ to the data yields $D_s^\perp  = (1.08 \pm 0.09\, \substack{+0.17 \\ -0.13})\, \hbar/m$, where the uncertainties are the statistical error from the fit  and the systematic error from the gradient calibration, respectively. This is a direct measurement of the time- and trap-averaged diffusivity that does not rely on any calibration other than that of the gradient.

In Fig.~\ref{fig:mag}B, we choose a constant gradient and vary $(T/T_F)_i$.  
Diffusivity is larger in hotter clouds, as both the typical velocity and the mean free path increase with temperature.
At lower temperature, we observe that $D_s^\perp$ does not continue to decrease, but appears to saturate. Careful examination of the demagnetization dynamics at our lowest initial temperatures (see insets to Fig.~\ref{fig:mag}B) also suggests an acceleration of demagnetization at later times.
An apparently time-dependent $D_s^\perp$ could be due to its polarization dependence, as
 is predicted below the so-called anisotropy temperature, where $D_s^\perp$ differs from $D_s^{||}$ \cite{Jeon:1989,Enss:2013ti}. It might also arise from spin-rotation effects \cite{Leggett:1970,Laloe:1982wc}. However, we find the deviations from Eq.~\ref{eq:tau} to be small, and we are unable to distinguish between these possibilities and other systematics. Within the probed range of temperature, the trap- and time-averaged $D_s^\perp$ is consistent with a quantum lower bound of $\simeq \hbar/m$.

Demagnetization transforms the system of $N$ particles in a single spin state to a mixture of two spin states, each with $N/2$ particles. The final Fermi energy of the trapped system $E_{F,f}$ therefore is reduced by a factor of $2^{1/3}$ compared to the initial $E_{F,i}$. Furthermore, demagnetization releases attractive interaction energy. Together these effects increase temperature \cite{SM}, so that each measurement of $D_s^{\perp}$ has to be understood as a time average over a range in temperatures. The intrinsic heating together with the initial polarization of the cloud ensures that the gas remains in the normal phase throughout the evolution \cite{Zwierlein:2006vf}.

The observation of suppressed spin transport indicates strong inter-particle scattering, but does not reveal how a thermodynamic interaction energy emerges. In a complementary set of measurements, we study the microscopic transformation of the gas by following the dynamical evolution of pair correlations that are enabled by demagnetization.  Instead of measuring $|M_\perp|$, we probe the gas with a pulse that couples $\up$ to $\probe$, an initially unoccupied internal state that interacts only weakly with $\up$ and $\dn$ \cite{SM}. The transfer rate to $\probe$ is measured as a function of the frequency detuning $\delta$ above the single-particle resonance. In a strongly interacting gas in equilibrium, the high-frequency tail of such a spectrum is known to be proportional to Tan's contact parameter $C=\int \mathbf{dr} \, {\cal{C}}(\mathbf{r})$ times $\delta^{-3/2}$ \cite{Tan:2008ey,Braaten:2008tc,Zhang:2009kq,Werner:2012um,Braaten2010,Randeria2010,Pieri:2009ty,Stewart:2010fy}.
The contact density ${\cal{C}}(\mathbf{r}) = \langle g^2 \psi^{\dagger}_{\uparrow}(\mathbf{r})\psi^{\dagger}_{\downarrow}(\mathbf{r})\psi_{\downarrow}(\mathbf{r})\psi_{\uparrow}(\mathbf{r})\rangle$ is a local measure of the pair correlation, i.e., the number of pairs of opposite spins at short distance, where $g$ is the coupling constant and $\psi_\sigma$ is the annihilation operator with spin $\sigma$. As is clear from its definition, $\cal{C}$ is also proportional to the local interaction energy. Although contact has been shown to relate various thermodynamic and many-body properties of a short-range interacting gas, it has so far been studied only in equilibrium and only with an unmagnetized gas \cite{Partridge:2005jh,Werner:2009vb,Zhang:2009kq,Stewart:2010fy,Vale2011}.

\begin{figure}[tb]\begin{center}
\includegraphics[width= 0.9 \columnwidth]{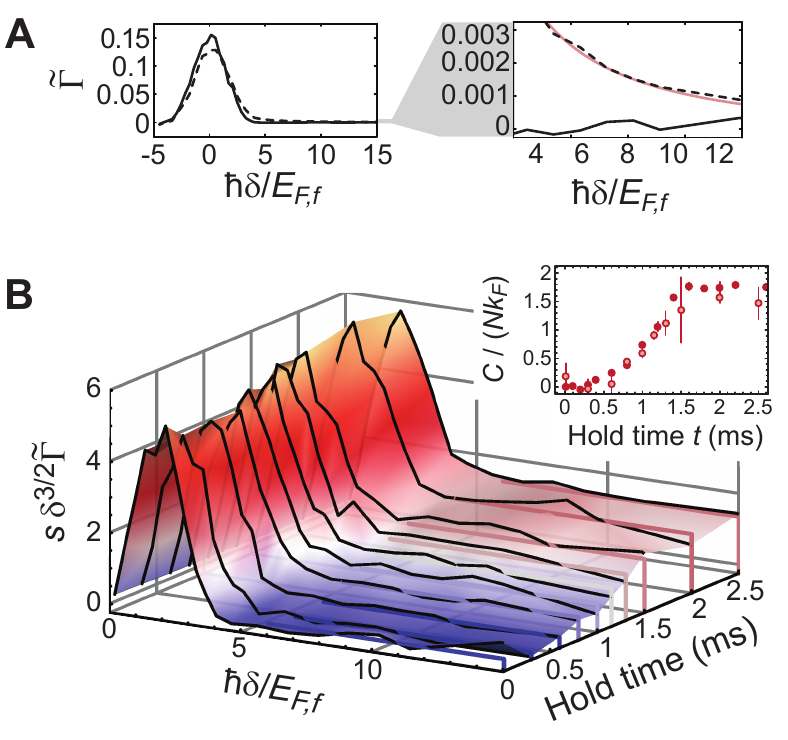}
\caption{{\bf Time-resolved spectroscopy to measure Tan's contact} at $B'=8.2(7)$\,G/cm and $(T/T_F)_i  \simeq 0.2$.
{\bf (A)} The normalized rate $\tilde \Gamma$ of atom transfer \cite{SM} from $\up$ to $\probe$ is shown after $10$\,$\mu$s (solid line) and $2.5$\,ms (dashed line), both without a spin-refocusing pulse. The red line in the right panel is a fit of a power-law $\propto \delta^{-3/2}$ to the high-frequency tail for $t = 2.5$\,ms.
{\bf (B)} The transfer rate $\tilde \Gamma$ rescaled with $s \delta^{3/2}$ versus hold time, where $s\equiv\pi^{2} (2 \hbar / E_{F,f})^{3/2}$, revealing a plateau for large positive detunings $\delta \gtrsim 4 E_{F,f}/\hbar$. 
The inset compares contact values extracted from these full spectra (open circles) and those measured at a fixed detuning $\delta/2 \pi= 125$\,kHz (filled circles). The latter method allows single-shot study of contact dynamics, and thus reduced statistical noise as seen here. The contact is normalized by the final Fermi momentum in the trap and the total number of atoms. Error bars in {B} are statistical.
\label{fig:spectrum}}
\end{center}\end{figure}

Figure~\ref{fig:spectrum}A shows that after a short hold time the spectrum exhibits only the single-particle peak, whereas after a longer hold time, the spectrum develops a high-frequency tail. Similar spectroscopic measurements starting from a polarized Fermi gas have shown
the emergence of mean-field shifts after decoherence \cite{Gupta:2003fz}. Here we study the high-frequency tail of the spectrum, finding that it has a $\delta^{-3/2}$ scaling at $\delta \gtrsim 4 E_{F,f}/\hbar$ for each hold time $t$, which indicates that pair correlations can be described with a contact parameter throughout the dynamics (Fig.~\ref{fig:spectrum}B).

\begin{figure}[tb]\begin{center}
\includegraphics[width= 0.9 \columnwidth]{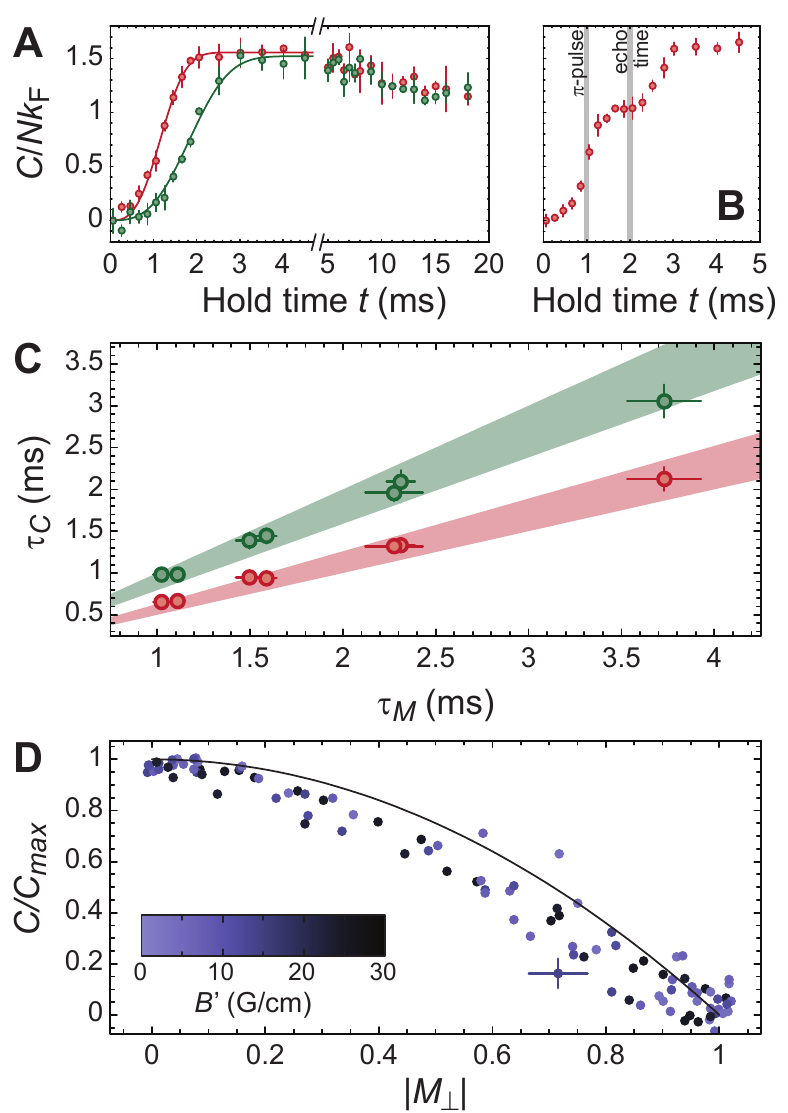}
\caption{{\bf Linking contact and magnetization.}
The time evolution of contact is studied using single-shot spectroscopy as illustrated by the inset of Fig.~\ref{fig:spectrum}B.
{\bf (A)} Contact versus hold time $t$ with (in green) and without (in red) a spin-refocusing $\pi$  pulse at $t/2$. Solid lines show a fit to $C_ {\rm max}(1 - \exp[-(t/\tau_C)^3])$.
{\bf (B)} Contact versus hold time $t$ with a $\pi$ pulse applied at $t_\pi = 1$\,ms. At the echo time, the spin spiral has untwisted and no longer drives a diffusive spin current. Error bars in A and B show statistical error across repeated measurements, and both data sets were taken in the same conditions as Fig.~\ref{fig:mag}A.
{\bf (C)} Contact growth time, $\tau_C$, versus demagnetization time, $\tau_M$ (colors as in A), measured at various gradients. Shaded areas are bounded by the limits of fully paired and uncorrelated spins (see text).
{\bf (D)} Normalized contact versus magnetization obtained by relating measurements of $|M_\perp(t)|$ and $C(t)$ for all sampled gradients $B'$. The solid line shows the behavior of uncorrelated spins. A typical statistical error bar is shown on a single point.  \label{fig:contactmag}}
\end{center}\end{figure}

Figure~\ref{fig:contactmag}A shows that, under various protocols, the contact starts at zero and grows in time towards a maximal value of $C_{\rm max}/(k_F N) = 1.53(4)$, where $\hbar k_F = \sqrt{2 m E_{F,f}}$ is the Fermi momentum in the final state of the trapped gas and $N$ is the total number of atoms. This is comparable to equilibrium values observed previously at $T/T_F \simeq 0.35$ in Ref.\ \citenum{Vale2011}, which lies between the initial and final temperatures of these data \cite{SM}.
At longer times ($t>5$\,ms), Fig.~\ref{fig:contactmag}A shows a slow reduction of contact, which is likely due to heating; however, in this work we focus on the short-time dynamics. A fit using an empirical rise function $f(t) = C_ {\rm max}(1 - \exp[-(t/\tau_C)^\eta])$ to the short-time data yields an exponent of $\eta = 3.6(3)$ with a spin-reversal and $\eta = 2.8(2)$ without a spin-reversal, reminiscent of the magnetization loss function $\propto\exp[-(t/\tau_M)^3]$. Further connection between contact and magnetization is demonstrated by Fig.~\ref{fig:contactmag}B, which traces the contact during a spin-reversal sequence: the rise of $C/(k_F N)$ is slowed by the refocusing pulse and plateaus at the spin-echo time, around which transverse spin diffusion is suppressed.

Figure~\ref{fig:contactmag}C compares $\tau_M$ and $\tau_C$, both with and without an echo. A linear relationship is found, which is surprising at first, since magnetization is a one-body vector observable and contact is a two-body scalar observable. The connection comes from the Pauli exclusion principle, which requires that if two particles are in the same location, as is required for a contact interaction, their spin state must be the antisymmetric spin-singlet state. For {uncorrelated} spin pairs, the probability to be in a spin-singlet state is $\rho_{ss} = (1 - |M_\perp|^2)/4$. Combining this assumption with the diffusion model for magnetization predicts $\tau_C=\tau_M/2^{1/3}$ and $\tau_C=\tau_M/2$ with and without an echo, respectively. The {maximum} singlet probability for a given magnetization is $\rho_{ss} = 1 - |M_\perp|$, and would instead give $\tau_C/\tau_M$ that is $2^{1/3}$ larger. Data in Fig.~\ref{fig:contactmag}C show an approximately linear relation whose slope is between these two limits.

Comparing the full range of measured values for normalized $C$ and $|M_\perp|$ at various times and gradients in Fig.~\ref{fig:contactmag}D also shows a functional form between the uncorrelated $C\sim1 - |M_\perp|^2$ and the fully paired $C\sim1 - |M_\perp|$. A calculation based on a large-$\cal{N}$ expansion \cite{Nikolic07,Veillette07,Enss12} predicts that $C(M_\perp)$ changes between these limiting behaviors as $T$ goes from $2T_c$ to $T_c$, where $T_c$ is the critical temperature for pair superfluidity \cite{SM}. Since a singlet pair has no net spin, the observation of enhanced $\rho_{ss}$ is also consistent with prior observations of reduced magnetic susceptibility due to strong attractive interactions in the normal state \cite{Sanner:2011ub,Nascimbene:2011fn}.

Alternatively, an apparent reduction in $C(M_\perp)$ might arise from a lag in the evolution of $C$ behind $|M_\perp|$. However we find no statistically significant dependence on gradient, which is evidence for a local equilibribration of $C$ on a faster time scale than the system-wide demagnetization. A true steady-state transport measurement, on the other hand, would suffer from an inhomogeneous magnetization due to imbalanced chemical potentials in the trap. Our dynamic measurement avoids this problem, since longitudinal spin transport is strongly suppressed on the millisecond time scale \cite{Hulet2011,Zwierlein2011}.

In conclusion, we have shown how a transversely spin-polarized Fermi gas decoheres and becomes strongly correlated at an interaction resonance. A diffusion constant of $\simeq \hbar /m$ challenges a quasiparticle-based understanding of transport by implying the necessity of maximally incoherent quasiparticles. A similar limit to the quasiparticle lifetime would explain the ubiquitous $T$-linear resistivity in metals \cite{Bruin:2013hc} and a quantum-limited shear viscosity \cite{Thomas2010}.

We thank B.~Braverman, I.~Kivlichan, L.~LeBlanc, and T.~Pfau for experimental assistance, D.~DeMille, T.~Enss, L.~Jiang, A.~Leggett, and A.~Paramekanti for discussions, A.~Aspect for manuscript comments, and M.~Ku and M. Zwierlein for sharing their unitary equation-of-state data. S.~T. acknowledges support from CQIQC. G.~E. acknowledges support from OGS. This work was supported by NSERC, CIFAR, the University of Hong Kong, and AFOSR under agreement number FA9550-13-1-0063.

\bibliography{bibV27}
\bibliographystyle{Science}

\setcounter{figure}{0}
\setcounter{equation}{0}
\renewcommand\thefigure{S\arabic{figure}}
\renewcommand\theequation{S\arabic{equation}}

\pagebreak
\centerline{\bf SUPPLEMENTARY MATERIAL}
\bigskip

\paragraph*{Sample preparation and thermometry.}
Fermionic, spin-polarized $^{40}$K atoms are cooled sympathetically with bosonic $^{87}$Rb atoms. Both species are initially trapped in a microfabricated magnetic trap, where $^{87}$Rb is evaporated directly. The final stage of cooling is performed in a crossed-beam optical dipole trap (ODT), with $^{40}$K atoms in the $\ket{F=9/2,m_F=-9/2}$ state and $^{87}$Rb atoms in the $\ket{F=1,m_F=1}$ state. At the end of cooling, residual $^{87}$Rb atoms are removed with a resonant light pulse. The ODT has a mean frequency $\bar\omega/2\pi = 486(15)$\,Hz, and an aspect ratio of 4:1:1. The Feshbach field $B_z$ and the gradient  $B' = \partial_z B_z$ are applied along a tight axis of the trap. The $\dn$, $\up$, and $\probe$ states in the main text refer to the high-field states adiabatically connected to the low-field $m_F=-9/2,-7/2,$ and $-5/2$ states of the $F=9/2$ hyperfine manifold of the electronic ground state.

The initial temperature is measured by imaging the atoms after expansion during time of flight and fitting a Fermi momentum distribution to the image. Since the gas is fully polarized in the $\ket{F=9/2,m_F=-9/2}$ state for this measurement, no interaction corrections are required. The data shown in Fig.~\ref{fig:mag}A and Fig.~\ref{fig:contactmag} was taken with $N = 3(1) \times 10^4$ atoms with an initial temperature of $330(30)$\,nK. 
The initial Fermi energy is then $E_{F,i} = h \times 27(3)$\,kHz, such that $(T/T_F)_i = 0.25(3)$, where $T_F \equiv E_\mathrm{F}/k_B$. The subscripts `$i$' and `$f$' indicate the values for the initial and final states, respectively.
The uncertainties stated for atom numbers and temperatures are a combination of statistical and calibration uncertainties. After complete demagnetization, the number of atoms per spin state is halved, to $N_\sigma = N/2$, and the Fermi energy drops by $2^{1/3}$ to $E_{F,f} =  h\times21(2)$\,kHz. This latter value, along with the corresponding $k_F = \sqrt{2 m E_{F,f}}/\hbar$, is used in the normalization of the contact spectra.

In order to tune $(T/T_F)_i$ for the measurements presented in Fig.~2B, we vary the loading and evaporation sequence, affecting both the absolute temperature of the gas and the total atom number. Spectroscopy data presented in Fig.~\ref{fig:spectrum} was taken with higher atom number, $N \simeq 5 \times 10^4$, to enhance the signal-to-noise ratio.

\begin{figure}[tb]
\centering
\includegraphics[width= 0.9 \columnwidth]{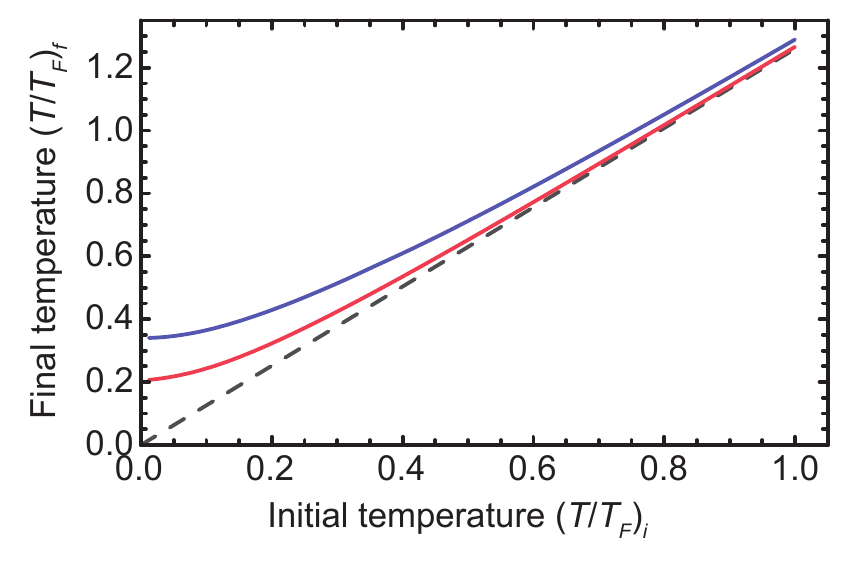}
\caption{Intrinsic heating during demagnetization. Final reduced temperature $(T/T_F)_f$ is shown as a function of initial reduced temperature $(T/T_F)_i$ for three different models: unitary mixture (blue solid line), ideal mixture (red solid line), and constant temperature (black dashed line). The constant-temperature line has a slope of $2^{1/3}$ due to the ratio of $E_{F,i}$ to $E_{F,f}$.
\label{fig:temprise} }
\end{figure}

Demagnetization releases energy and increases entropy, changing the temperature. We refer to this effect as `intrinsic heating'. Figure~\ref{fig:temprise} shows a calculation of the final reduced temperature $(T/T_F)_f$ in terms of the initial reduced temperature $(T/T_F)_i$. The calculation assumes conserved total energy and particle number, since the trapped gas is isolated. The red line shows the temperature rise that would occur in an ideal gas, as discussed in Ref.~\citenum{Ragan:2002tx}. However demagnetization at unitarity also releases an attractive interaction energy, further increasing temperature (blue line in Fig.~\ref{fig:temprise}). This prediction is calculated using the measured density equation of state \cite{Ku:2012ue}. At low temperature, the effect is strong: a $(T/T_F)_i=0$ cloud will heat to $(T/T_F)_f \approx 0.35$, and at our lowest $(T/T_F)_i \approx 0.15$, the final reduced temperature is $(T/T_F)_f \approx 0.40$.  At high temperature, the intrinsic heating effect is small, and these three lines shown in Fig.~\ref{fig:temprise} approach each other.

\paragraph*{Magnetic field control.}
We control the magnetic field and its gradients through a combination of magnetic field coils and micro-fabricated wires on an atom chip located about $200$\,$\mu$m from the atoms. We tune the field to be $|{\bf B}| = B_z = 202.10(2)$\,G, at which the two states $\up$ and $\dn$ undergo a Feshbach resonance. Recent measurements \cite{Gaebler:2010kb,Ludewig:2012vv} differ by 0.1\,G on the location of the resonance, giving a systematic uncertainty of $\pm0.1$ in $(k_F a)^{-1}$. The field is calibrated by measuring the $\dn$ to $\up$ transition frequency, which we find to be  $2 \pi \times 44.817(4)$\,MHz, and the uncertainty represents day-to-day fluctuation of the 1-kHz-wide spectral line. We control the field gradients $\partial_z B_z$ and $\partial_y B_z$ by adjusting the sum and the difference of small currents through parallel chip wires near the atoms (typically setting $\partial_y B_z=0$). We directly calibrate the gradients by repeating spectroscopy measurements on a cloud translated by piezo-actuated mirrors on the trapping beams. The cloud position is measured using two orthogonal imaging systems, whose magnification is calibrated by a dropped $^{87}$Rb cloud in a magnetic-field-insensitive state. The systematic gradient uncertainty given in the main text is dominated by the magnification uncertainty.

\paragraph*{Magnetization measurement.} 
Magnetization decay is measured using a $\pi/2$\,--\,$\pi$\,--\,$\pi/2$  sequence. The first and second pulses have the same phase, but the final pulse has a variable relative phase $\phi$. Varying $\phi$ reveals the magnitude of the transverse magnetization, $|M_\perp|$, in the visibility of the oscillation in relative population.  We are sensitive to the relative frequency between the drive $\omega$ and the atomic resonance $\omega_0$, with a precision of roughly $1/t$, where $t$ is the hold time. For $t\gtrsim1$\,ms, we find that our field stability (roughly 1\,kHz, or a few parts in $10^5$) is insufficient to preserve a reproducible relative phase throughout the sequence, resulting in a randomized phase for long hold times. To avoid an underestimation of the magnetization coherence time, we use the standard deviation of $N_\uparrow/N$ as a measure of $|M_\perp|$. Each measurement of $|M_\perp|$ at a single hold time consists of 12 shots, ensuring sufficient sampling of the randomized phase. We find no systematic difference between $\tau_M$ measured using rms and extremal values of the $M_z$ distribution.

\paragraph*{State-selective imaging.} 
Our imaging scheme allows us to simultaneously count the populations of atoms in states $\dn$ and $\probe$, leaving atoms in state $\up$ invisible. This is achieved with a Stern-Gerlach pulse to separate the trapped spin states, radiofrequency (rf) state manipulation during time-of-flight in a gradient, and finally,  resonant absorption imaging on the $\ket{F=9/2,m_F=-9/2}$ to $\ket{F=11/2,m_F=-11/2}$ cycling transition. Imaging occurs after jumping the magnetic field to 209\,G, the zero crossing of the s-wave scattering resonance, in order to minimize interaction effects during time of flight. In the magnetization measurements, we include an additional step transferring atoms in state $\up$ to $\probe$ with an adiabatic rapid passage prior to imaging.

\paragraph*{Normalization and rescaling of rf spectra.}
The rate $\Gamma$ of atom transfer from state $\up$ to the weakly
interacting probe state $\probe$ obeys the sum rule $\int
\Gamma(\omega)d\omega = \frac{1}{2} \pi \Omega_R^2 N_{\uparrow}$, where $\Omega_R$ is the
Rabi frequency of the applied rf pulse.
In the presence of strong interactions, $\Gamma(\omega)$ scales as $\Omega_R^2
/ (8 \pi \sqrt{m/\hbar}\, \delta^{3/2}) C$ for large positive detunings
$\delta = \omega - \omega_0$, where $\omega_0$ is the single-particle
resonance frequency. We express the detuning in units of the Fermi energy,
$\Delta = \hbar \delta/E_{F,f}$, where $E_{F,f}$ is calculated for a balanced mixture
with $N_\uparrow=N_\downarrow=N/2$. We furthermore
introduce the normalized transfer rate $\tilde \Gamma(\Delta) = E_{F,f}/(\hbar \pi
\Omega_R^2 N_\uparrow) \Gamma$ such that $\int \tilde \Gamma(\Delta)d\Delta =
1/2$. This rescaled transfer rate $\tilde \Gamma$ is plotted in Figure 3A versus $\Delta$.
Its asymptotic behavior takes the form
\begin{equation}
	\tilde \Gamma(\Delta) \to \frac{1}{2^{3/2}\pi^2 \Delta^{3/2}}  \frac{C}{k_F
N}\,,
\end{equation}
so that $\tilde \Gamma(\Delta)/(2^{3/2}\pi^2 \Delta^{3/2})$, as plotted in Fig.~\ref{fig:spectrum}B, reveals ${C}/(k_F N)$, the contact per particle in units of $k_F$, at $\Delta \gg 1$ \cite{Braaten2010,Randeria2010,Pieri:2009ty,Stewart:2010fy}.

The normalized transfer rate in Fig.~\ref{fig:spectrum}A is $\tilde \Gamma$, and the rescaled transfer rate in Fig.~\ref{fig:spectrum}B is $\tilde \Gamma$ times $ 2^{3/2}\pi^2 \Delta^{3/2}$.

In our experiments, we measure the fraction of atoms $N_{p}/N_\uparrow$
transferred to the probe state by a Blackman pulse of length $\Delta t$ and
with mean Rabi frequency $\Omega_R$. We choose the rf power and pulse duration
such that we probe the transition in the linear regime where
$\Gamma(\delta) = N_{p}(\delta)/\Delta t$. Using the asymptotic
behavior stated above, we obtain the contact from the fraction of atoms
transferred, measured at a single detuning $\delta$, via
\begin{equation}
	\frac{C}{k_F N} = \frac{4\pi}{\Omega_R^2 \Delta t}
\delta^{3/2} \sqrt{\frac{m}{\hbar}} \frac{1}{k_F}
\frac{N_p}{N_\uparrow}\,,
\end{equation}
where the Fermi wave number $k_F$ is calculated using the measured value for
$N_\uparrow=N/2$.

\paragraph*{Magnetization dependence of the contact in a normal, spin-polarized Fermi gas.}
Here we consider the magnetization dependence of the contact density $\mathcal{C}$ for a two-component gas of fermions described by the Hamiltonian
\begin{eqnarray}
{\cal{H}}&=&\sum_\sigma\int d^3{\bf r}\;\psi_\sigma^\dagger({\bf r})(-\frac{\hbar^2\nabla^2}{2m}-\mu_\sigma)\psi_\sigma({\bf r}) \nonumber\\
&&\phantom{\sum_\sigma\int d^3{\bf r}}+\;g \psi_\uparrow^\dagger({\bf r})\psi_\downarrow^\dagger({\bf r})\psi_\downarrow({\bf r})\psi_\uparrow({\bf r}).
\label{H}\end{eqnarray}
The chemical potentials $\mu_\sigma$ are fixed by the constraint that the total density $n$ and the (normalized) magnetization are kept constant. $g \equiv [m/4\pi\hbar^2 a - \sum_{|\mathbf{k}|\leq \Lambda}m/(\hbar|\mathbf{k}|)^2]^{-1}$ is the coupling constant~\cite{deMelo93}, regularized by the ultraviolet momentum cutoff $\Lambda$, and $a$ is the s-wave scattering length.   In our experiment, ${\bf M}$ lies in the transverse plane.  However, the interaction term is isotropic ($s$-wave scattering) and as a result, the contact density $\mathcal{C}$, being a scalar quantity, only depends on the magnitude of $|{\bf M}|$.  Thus, in the discussion below, we take ${\bf M}= {M}\hat{z}$, without affecting the final conclusions. In terms of the density of spin states, $M=(n_\uparrow-n_\downarrow)/n$.

A useful definition of the contact that makes clear its connection with interactions is~\cite{Braaten:2008tc}
\begin{equation} {\cal{C}} = \langle g^2 \psi^{\dagger}_{\uparrow}(\mathbf{r})\psi^{\dagger}_{\downarrow}(\mathbf{r})\psi_{\downarrow}(\mathbf{r})\psi_{\uparrow}(\mathbf{r})\rangle. \label{Cdef} \end{equation}
This allows us to make the following observation: at sufficiently high temperatures, even though the interactions are strong in a gas near unitarity ($a/\lambda_T\to \infty$; $\lambda_T$ is the thermal de Broglie wavelength), many-body correlations are weak ($\lambda_T\ll n^{-1/3}$) and Eq.~\ref{Cdef} can be factorized, 
\begin{equation}
\mathcal{C}(M)/\mathcal{C}_0=1-{M}^2,
\label{CM}
\end{equation}
where $\mathcal{C}_0\equiv \mathcal{C}({M}=0)$ is the value of contact density at zero magnetization.

At lower temperatures, $T\lesssim T_F$, $\lambda_T$ is greater than the mean distance $n^{-1/3}$ between atoms, and quantum many-body correlations become significant when the scattering length is large.  In particular, pairing correlations associated with the formation of incoherent pairs above $T_c$ are expected to modify the quadratic dependence of ${\cal{C}}({M})/{\cal{C}}_0$ shown in Eq.~\ref{CM}.   We note, for instance, that for an ideal gas of dimer molecules -- representing the BEC limit with maximal pairing correlations -- with binding energy $E_b = -\hbar^2/ma^2$, the contact is ${\cal{C}}  = (4\pi n/a)(1-{M})$. Thus,
\begin{equation} {\cal{C}}({M})/{\cal{C}}_0 = 1-{M}.\label{CMmol}\end{equation}
It seems reasonable to expect that the incoherent pairs in a unitary Fermi gas in the regime $T_c\lesssim T\lesssim T_F$ would give rise to a ${\cal{C}}({M})/{\cal{C}}_0$ curve that is intermediate between the two extremes, Eqs.~\ref{CM} and \ref{CMmol}. 

One of the few techniques for treating such pairing effects systematically in the strong-interaction regime is the large-$\mathcal{N}$ expansion~\cite{Nikolic07,Veillette07,Enss12}.  This approach artificially generalizes the Sp(2) quantum field theory  Eq.~\ref{H} of the dilute Fermi gas to an Sp(2$\mathcal{N}$) theory with $\mathcal{N}$ ``flavors'' of spin $\uparrow$ and $\downarrow$ fermions interacting via an Sp(2$\mathcal{N}$)-invariant attractive interaction. The $\mathcal{N} = \infty$ limit of this theory corresponds to the high-temperature ``classical theory" in which Eq.~\ref{CM} is strictly satisfied; many-body quantum correlations are brought in via corrections in powers of $1/\mathcal{N}$ and are expected to lead to deviations from Eq.~\ref{CM} at low temperatures.    

\begin{widetext}
To implement the large-$\mathcal{N}$ expansion, we write the grand canonical thermodynamic potential density $\Omega$ as a functional integral over the Bose pairing field $\phi$~\cite{deMelo93,Veillette07}, where $\beta\equiv (k_BT)^{-1}$: 
\begin{equation} \Omega = -\beta^{-1} \ln \int {\cal{D}}[\phi^*, \phi]\exp \left\{ \mathcal{N}\int^{\beta}_0\!\!d\tau\!\!\int d^3{\mathbf{r}}\frac{|\phi(\mathbf{r},\tau)|^2}{g} + \mathcal{N}\mathrm{Tr}\ln [-\mathbf{G}^{-1}]\right\}.\label{Omega}\end{equation}
Here the inverse matrix Nambu-Gor'kov Green's $\mathbf{G}^{-1}$ function is [$x\equiv (\mathbf{r},\tau)$]
\begin{equation} \mathbf{G}^{-1}(x,x')\equiv \left(\begin{array}{cc} -\hbar\partial_{\tau} + \frac{\hbar^2\nabla^2}{2m}+\mu_{\uparrow} & \phi(x)\\
\phi^*(x)& -\hbar \partial_{\tau}-\frac{\hbar^2\nabla^2}{2m}-\mu_{\downarrow}\end{array}\right)\delta(x-x'),\end{equation}
and the trace in Eq.~\ref{Omega} is over space $\mathbf{r}$, imaginary time $\tau$, as well as matrix indices. 
\end{widetext}

\begin{figure}[b]
\centering
\includegraphics[width=0.9 \columnwidth]{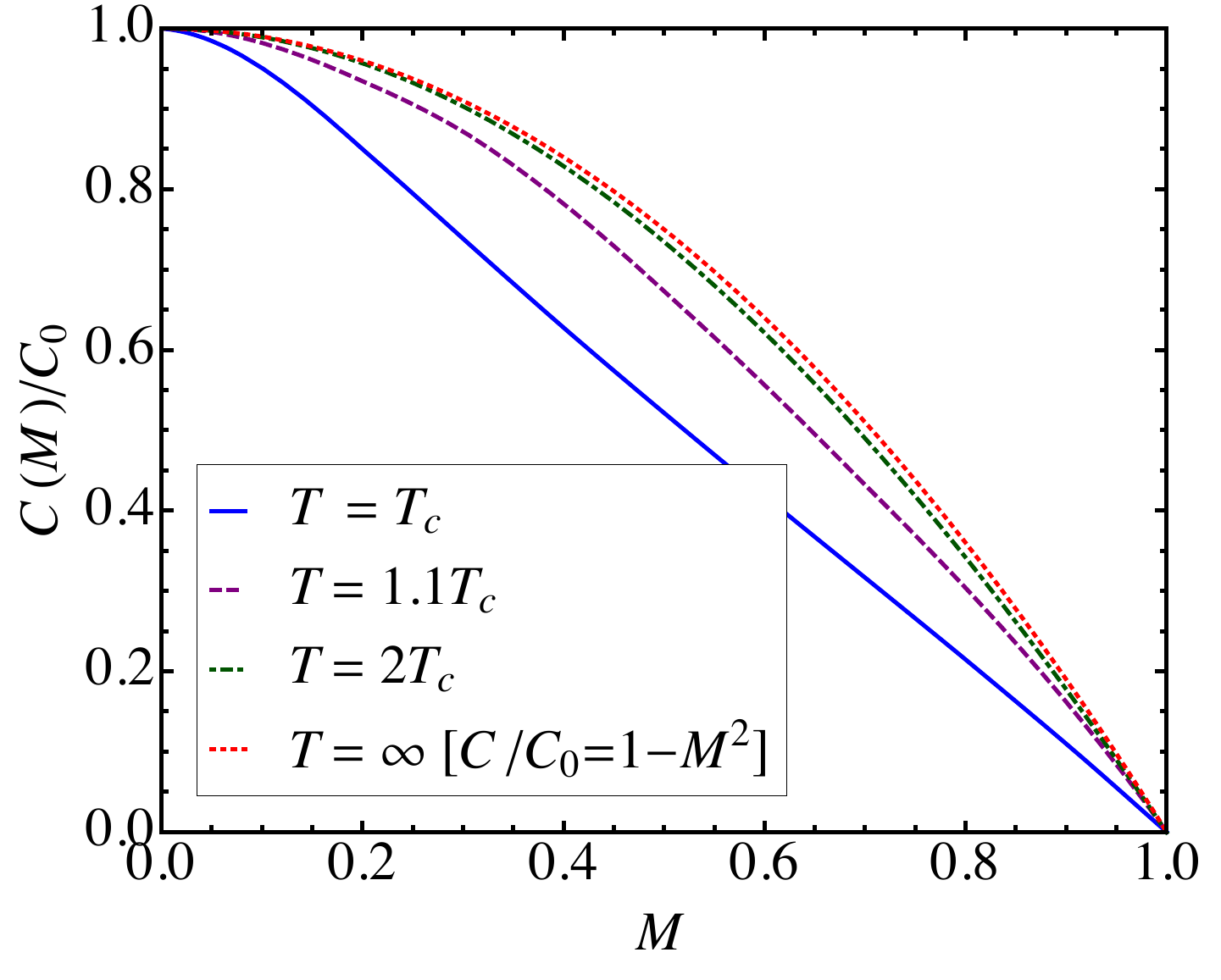}
\caption{Magnetization dependence of the contact in a normal unitary Fermi gas to leading order in a large-$\mathcal{N}$ calculation at various temperatures, scaled by ${\cal{C}}(M=0)$.  Deviations from the high-$T$ asymptote ${\cal{C}}(M)/{\cal{C}}_0 = 1-M^2$ in the region above $T_c$ signal the onset of strong pairing correlations.} 
\label{CM-fig}
\end{figure}

Using standard thermodynamic identities, one can rewrite Tan's  ``adiabatic relation"~\cite{Tan:2008eg} ${\cal{C}} = -(4\pi m/\hbar^2) (\partial{\cal{E}}/\partial a^{-1})_S$, where ${\cal{E}}$ is the energy density and $S$ is the entropy, as 
\begin{equation} {\cal{C}} = -\frac{m^2}{\hbar^4} \left(\frac{\partial \Omega}{\partial g^{-1}}\right)_{\!\mu_{\sigma},T}.\end{equation}
Applying this to Eq.~\ref{Omega} (with $\mathcal{N}=1$), it is immediately apparent that the contact is just the expectation value of the square of the amplitude of the Bose pairing field (see also Chs.~6 and 8 by Braaten and Sachdev, respectively, in Ref.~\citenum{Zwerger:2012}): ${\cal{C}} = (m^2/\hbar^4\beta)\int^{\beta }_0d\tau \int d^3\mathbf{r} \langle |\phi(\mathbf{r},\tau)|^2\rangle$.   Formally, this is an exact identity.  To make progress, however, we need to expand the argument of the exponential in Eq.~\ref{Omega} to order $1/\mathcal{N}$ (i.e., Gaussian with respect to $\phi$).  In the normal state ($\langle \phi\rangle=0$) above $T_c$, Fourier transforming to momentum $\mathbf{q}$ and Matsubara frequency $\nu_m$ space, the ${\cal{O}}(1/\mathcal{N})$ contribution to the thermodynamic potential is~\cite{deMelo93} 
\begin{equation}
\Omega =  -\beta^{-1}\ln \int\!\!{\cal{D}}[\phi^*,\phi]\exp\left[\mathcal{N}\sum_{\mathbf{q},\nu_m}\Gamma^{-1}(\mathbf{q},i\nu_m)|\phi(\mathbf{q},i\nu_m)|^2\right]
\end{equation}
and hence (there is no ${\cal{O}}(1/\mathcal{N})^0$ contribution in the normal state since $\langle \phi\rangle = 0$)~\cite{Enss12},
\begin{equation} {\cal{C}} = \frac{m^2 }{\hbar^4\beta \mathcal{N}}\sum_{\mathbf{q},\nu_m}\Gamma(\mathbf{q},i\nu_m) + {\cal{O}}(1/\mathcal{N})^2. \label{CN}\end{equation}
Here the two-particle vertex function $\Gamma$ is defined by its inverse
\begin{equation} \Gamma^{-1}(\mathbf{q},i\nu_m) = \frac{1}{g}-\sum_{\mathbf{k}}\frac{1-f(\xi_{\mathbf{k}\uparrow})-f(\xi_{\mathbf{k}\downarrow})}{\hbar i\nu_m - \xi_{\mathbf{k}\uparrow}-\xi_{\mathbf{k}+\mathbf{q}\downarrow}}.\label{vertex}\end{equation}
 $f(\xi_{\mathbf{k}\sigma})$ is the Fermi-Dirac distribution for spin $\sigma$ and $\xi_{\mathbf{k}\sigma}\equiv \hbar^2|\mathbf{k}|^2/2m-\mu_{\sigma}$.  Since Eq.~\ref{CN} is already order $1/\mathcal{N}$, the chemical potentials that enter the two-particle vertex function can be obtained from the ${\cal{O}}(1/\mathcal{N})^0$ mean-field number equations $n = \sum_{\mathbf{k}}[f(\xi_{\mathbf{k}\uparrow})+f(\xi_{\mathbf{k}\downarrow})]$ and ${M} = \sum_{\mathbf{k}}[f(\xi_{\mathbf{k}\uparrow})-f(\xi_{\mathbf{k}\downarrow})]/n$.  

Solving Eqs.~\ref{CN} and \ref{vertex} in conjunction with the mean-field number equations at unitarity gives the contact as a function of magnetization shown in Fig.~\ref{CM-fig} for several values of $T/T_c$.  To leading order in $1/\mathcal{N}$, the critical temperature where the pairing susceptibility diverges, $\Gamma(\mathbf{0},0)=\infty$, is the mean-field value, $T_c\sim 0.5T_F$.  At high-temperatures $T\gg T_c$, where the fugacities $z\equiv \exp(\beta \mu_{\sigma})$ are small, it is straightforward to show that Eqs.~\ref{CN} and \ref{vertex} reduce to
${\cal{C}}({M}, \beta \mu_{\sigma}\ll 0) = (16\pi z_{\uparrow}z_{\downarrow})/\lambda^4_T$.  Using the high-$T$ limits $n\lambda^3_T \to  z_{\uparrow}+z_{\downarrow}$ and ${M}n\lambda^3_T \to  z_{\uparrow}-z_{\downarrow}$ of the number equations, this gives
\begin{equation} {\cal{C}}({M}, \beta \mu_{\sigma}\ll 0) = 4\pi n^2\lambda^2_T(1-{M}^2),\end{equation}
in agreement with the expected high-$T$ asymptote Eq.~\ref{CM} as well as the leading order virial expansion value of the contact at zero magnetization~\cite{Yu09}.  From Fig.~\ref{CM-fig}, we see that the contact saturates to this asymptotic $1-{M}^2$ behavior already at only moderately high temperatures, $T\sim 2T_c$. Quantum corrections are only evident at lower temperatures, where strong enhancement of the pairing susceptibility leads to a suppression of ${\cal{C}}({M})/{\cal{C}}_0$, suggestive of the pairing behavior responsible for Eq.~\ref{CMmol}.  This effect is seen as well in Fig.~4D in the main text although our simple calculation here likely underestimates the range of temperatures above $T_c$ where this effect is evident.

%
%
%
%
%
%
%

\end{document}